\pdfoutput=1
\documentclass[twocolumn,english,aps,prb]{revtex4}
\usepackage[T1]{fontenc}
\usepackage[latin9]{inputenc}
\usepackage{bm}
\usepackage{amsmath}
\usepackage{amssymb}
\usepackage{graphicx}
\usepackage{esint}

\makeatletter
\@ifundefined{textcolor}{}
{%
 \definecolor{BLACK}{gray}{0}
 \definecolor{WHITE}{gray}{1}
 \definecolor{RED}{rgb}{1,0,0}
 \definecolor{GREEN}{rgb}{0,1,0}
 \definecolor{BLUE}{rgb}{0,0,1}
 \definecolor{CYAN}{cmyk}{1,0,0,0}
 \definecolor{MAGENTA}{cmyk}{0,1,0,0}
 \definecolor{YELLOW}{cmyk}{0,0,1,0}
 }

\@ifundefined{definecolor}
 {\usepackage{color}}{}
\@ifundefined{definecolor}{\usepackage{color}}{}
\@ifundefined{definecolor}{\@ifundefined{definecolor}
 {\usepackage{color}}{}
}{}

\usepackage{babel}

\makeatother

\usepackage{babel}
\begin{document}

\title{Theory of inelastic light scattering in spin-1 systems: resonant
regimes and detection of quadrupolar order}

\author{F. Michaud$^{1}$, F. Vernay$^{2}$, and F. Mila$^{1}$}

\affiliation{$^{1}$CTMC, Ecole Polytechnique Fédérale de Lausanne, CH-1015 Lausanne,
Switzerland\\
 $^{2}$LAMPS, Université de Perpignan Via Domitia, F-66860 Perpignan,
France}

\date{\today}
\begin{abstract}
Motivated by the lack of an obvious spectroscopic probe to investigate
non-conventional order such as quadrupolar orders in spin \emph{$S>\frac{1}{2}$}
systems, we present a theoretical approach to inelastic light scattering
for spin-1 quantum magnets in the context of a two-band Hubbard model.
In contrast to the $S=\frac{1}{2}$ case, where the only type of
local excited state is a doubly occupied state of energy $U$, several local
excited states with occupation up to 4 electrons are present. As a consequence,
we show that two distinct resonating scattering regimes can be accessed depending
on the incident photon energy. For $\hbar\omega_{in}\lesssim U$, the
standard Loudon-Fleury operator remains the leading term of the expansion
as in the spin-$\frac{1}{2}$ case. For $\hbar\omega_{in}\lesssim4U$,
a second resonant regime 
is found with a leading term that takes the form of a biquadratic coupling
$\sim\left({\bf S}_{i}\cdot{\bf S}_{j}\right)^{2}$. Consequences for the
Raman spectra of S=1 magnets with magnetic or quadrupolar order are
discussed. Raman scattering appears to be a powerful probe
of quadrupolar order.
\end{abstract}
\maketitle

\section{Introduction}

The theoretical quest for exotic phases of magnetic matter over the
last few decades has opened the way for a systematic investigation
of models by various analytical and numerical means and has led to
the conceptual understanding of different states. This is for instance the
case of resonating valence bond states which have been investigated
in the context of high-Tc superconductors, \citep{Anderson87}
frustrated spin systems\citep{Mila98} and quantum
dimer models.\citep{Rokhsar,Moessner01} Yet, suggesting a technique
to unravel the experimental fingerprints of an exotic phase often
remains a challenge: for instance, in the spin liquid case, it is
only recently that theoretical proposals have been made to detect
this state on the kagome lattice with Raman scattering\citep{Lhuillier08,Lee10}
and that experiments have been carried out.\citep{Lemmens-kagome}

Raman scattering was discovered in the beginning of the twentieth
century. Loudon and Fleury showed how to use it to detect magnetic
excitations in the beginning of the sixties.\citep{Fleury68} It was
used for instance in the late eighties to estimate the exchange in
the cuprates;\citep{Lyons} more recently the investigation of magnetic
properties with this spectroscopic technique
in various contexts like frustrated systems\citep{Lee10,Lhuillier08,Perkins,Vernay}
or iron pnictides\citep{Cheng-Chien} has been a very active field of research.
In fact, inelastic light scattering
takes advantage of the fact that the response is in essence linked to many-body
physics and that the photon polarization enables to collect valuable
insights on correlated electron systems.\citep{Devereaux07,Lemmens_Review} Following
this route, we present here inelastic light scattering as a natural
probe to characterize the order and the low-energy excitations in
spin $S=1$ models.

The pecularity of quantum magnets with $S>1/2$ is that they can
break SU(2) symmetry with a local order parameter which is non
magnetic. The simplest example is provided by quadrupolar order
for $S=1$ systems, where the local order parameter is not the
spin but a rank-2 tensor. This has been extensively studied in the context of
the bilinear-biquadratic model defined by the Hamiltonian:
\begin{equation}
\mathcal{H}_{eff}=J\sum_{\langle i,j\rangle}\left[\cos\theta\left({\bf S}_{i}\cdot{\bf S}_{j}\right)+\sin\theta\left({\bf S}_{i}\cdot{\bf S}_{j}\right)^{2}\right]\label{eq:bilinear-biquadratic}
\end{equation}
Ferro- and antiferroquadrupolar phases have been identified both on the
triangular\cite{Arikawa06,Mila06} and square\cite{Kawashima,Toth10} lattices for sufficiently
large biquadratic interactions, and the compound NiGa$_{2}$S$_{4}$, where Ni$^{2+}$ ions
form a triangular lattice of spins 1, has been suggested to exhibit some kind of quadrupolar order\cite{Nakatsuji,Nakatsuji07,Arikawa06,Mila06}. Possible mechanisms to produce large biquadratic
interactions include spin-lattice coupling\cite{Kittel_magnetostriction_PRB1960} and situations with quasi-orbital degeneracy\citep{Mila00}.

A direct observation of quadrupolar order remains a challenge however.\citep{Arikawa06}
Quadrupolar states being non-magnetic, conventional experimental techniques such
as neutron scattering are insensitive to quadrupolar order.
By contrast, we show in the present work that, since light naturally couples
to the charge, inelastic light scattering offers an alternative to
investigate non-magnetic states.

The paper is organized as follows: in a first section we present the
microscopic model and the derivation of the effective magnetic
light-scattering operator. We discuss the form of this effective $S=1$
operator compared to the more conventional $S=\frac{1}{2}$ case.
We show that two different resonant regimes are accessible, depending
on the incoming photon energy and we discuss the different polarization
geometries relevant to our problematic. This leads to a section
in which the expected spectra for the different phases are displayed,
followed by a section devoted to a discussion concerning potential
experimental checks, and by a brief conclusion.

\section{Effective light scattering operator}

\subsection{Microscopic parameters\label{sub:Microscopic-parameters}}

From the point of view of purely atomic physics, spin-1 states can
be achieved in the case of transition metal ions, for instance Ni$^{2+}$
($3d^{8}$) in a cubic environment that leads to the standard $t_{2g}-e_{g}$
splitting of the orbital $d$-shell. The intra-atomic electron-electron
interaction of the partially filled shell can then be reformulated
as a Hund's coupling which favors states maximizing the spin.

As we want to keep a very general Hubbard model, we will consider
here a lattice of such sites: two degenerate orbitals at half-filling,
\emph{i.e.} 2 electrons per site, with nearest-neighbor hopping and
on-site interactions that include inter- and intra-orbital coupling.
This two-band Hubbard model is described by the following Hamiltonian:

\begin{eqnarray}
\mathcal{H}_{Hb}=\sum_{i,j}\sum_{m,m'=a,b}t_{m,m'}^{ij}c_{im\sigma}^{\dagger}c_{jm'\sigma} &  & {\rm }\label{eq:microscopic}\\
+\frac{1}{2}\sum_{m,m'}\sum_{\sigma\sigma'}U_{mm'}n_{jm\sigma}n_{jm'\sigma'} &  & {\rm }\nonumber \\
+\frac{1}{2}\sum_{m\neq m'}\sum_{\sigma\neq\sigma'}\{J_{H}n_{jm\sigma}n_{jm'\sigma} &  & {\rm }\nonumber \\
+J_{H}c_{im\sigma}^{\dagger}c_{im\sigma'}c_{im'\sigma'}^{\dagger}c_{im'\sigma} &  & {\rm }\nonumber \\
+J_{P}c_{im'\sigma'}^{\dagger}c_{im'\sigma}^{\dagger}c_{im\sigma'}c_{im\sigma}\} &  & {\rm }\nonumber
\end{eqnarray}
 where $i,j$ are the site indices, $m,m'$ refer to the orbitals
and $\sigma$ to the electronic spin. The hopping parameters between
two neighboring orbitals is $t_{m,m'}$, the on-site Coulomb repulsion
is denoted by $U_{mm'}$, $J_{H}$ represents the Hund's coupling
and $J_{P}$ the pair hopping amplitude.
For the sake of clarity, we restrict ourselves here to a square lattice,
although the extension to other lattices is straightforward. Furthermore,
we assume that the additional relations, typical of cubic symmetry, are
satisfied, namely:
$U_{aa}=U_{bb}$, $U=U_{aa}-2J_{H}$ and $J_{H}=2J_{P}$.

In the Mott insulator regime, at half-filling and for $U_{mm'}\gg t_{m,m'}$,
the electrons are localized and Hund's coupling favors triplet states
on each site. The relevant degree of freedom is then a spin $S=1$.
At second order in perturbation theory, the resulting effective spin-1
Hamiltonian is the standard Heisenberg
model:
\[
\mathcal{H}_{eff}^{(2)}=J_{Heis}^{(2)}\sum_{\langle i,j\rangle}{\bf S}_{i}\cdot{\bf S}_{j}
\]
with
\[
J_{Heis}^{(2)}=\frac{t_{aa}^{2}+2t_{ab}^{2}+t_{bb}^{2}}{U+2J_{H}}.
\]
As in the case of $S=1/2$ effective models for single-band Hubbard,\citep{MacDonald88,Gingras05}
pushing the perturbation to fourth order leads to the emergence of
additional terms such as 4-site terms\citep{Bastardis07} and biquadratic interactions.

\subsection{Derivation of the effective operator}

\subsubsection{General method\label{sub:General-method}}

Inelastic light scattering techniques like Raman are photon-in photon-out
techniques where the incident photon couples to the charge. Shastry
and Shraiman gave a microscopic description of the process in Ref.~\onlinecite{shastry90}.
The procedure is however given in a more pedagogical way in Ref.~\onlinecite{Lee10}.
As we follow very closely the latter, we will not explain in great
detail the derivation but just sketch the main ideas. The photon-electron
interaction is introduced via the Peierls substitution:
\[
c_{ix\sigma}\to c_{ix\sigma}\exp\left[-i\frac{e}{\hbar c}\int_{-\infty}^{r_{i}}{\bf A}\cdot{\bf dl}\right];
\]
 where ${\bf A}$ is the photon vector potential.

For incoming photon wave-lengths much larger than the lattice spacing,
it can easily be shown that the coupling of the microscopic Hamiltonian
to the photon affects only the kinetic terms and generates a current
that depends on the incident ($\bm{e_{in}}$) and scattered ($\bm{e_{out}}$)
polarizations. After second quantization of ${\bf A}$, $a$ and $a^{\dagger}$
being respectively the photon creation and annihilation operators,
the current term reads:
\[
\begin{array}{ccc}
\mathcal{H}_{C} & = & i\frac{e}{\hbar c}{\displaystyle \sum_{\left\langle i,j\right\rangle }}{\displaystyle \sum_{\begin{array}{c}
m,m'\end{array}}}t_{m,m'}^{ij}c_{im\sigma}^{\dagger}c_{jm'\sigma}\\
\\
 & \times & {\displaystyle \sum_{k_{in},k_{out}}}\left(g_{in}\boldsymbol{e_{in}}a_{k_{in}}+g_{out}\boldsymbol{e_{out}}a_{k_{out}}^{\dagger}\right)\boldsymbol{\cdot e_{i\rightarrow j}}
\end{array}
\]
 with $g_{in,out}=\sqrt{hc^{2}/\omega_{k_{in,out}}V}$, where
$V$ stands for the volume, and $\bm{e_{i\to j}}$ is the vector connecting
site $i$ and site $j$.

Hence, collecting all the parts of the Hamiltonian describing the
electronic system coupled to the light we have: $\mathcal{H}=\mathcal{H}_{Hb}+\mathcal{H}_{C}+\mathcal{H}_{\gamma}$,
where $\mathcal{H}_{\gamma}$ represents the purely photonic part
of the Hamiltonian.

In the limit $U\gg t$, the electrons are localized and the low-energy
spectrum of the system can be described by a spin Hamiltonian. Furthermore,
in the limit $\left|U-\omega_{in}\right|\lesssim t$, $\mathcal{H}_{c}$
can be treated as a perturbation and one can derive an effective magnetic
Raman operator in the spin sector. Yet, if one is interested to have
an effective operator beyond the standard limit $\left|U-\omega_{in}\right|\ll t$,
one has to push the perturbation theory to fourth order. It is worth
pointing out that this effective description conserves the quantum
numbers of the original Hamiltonian and scattering operator: for instance
since $\mathcal{H}_{Hb}$ and $\mathcal{H}_{C}$ conserve the total
spin, so will the effective Raman operator and thus we recover the
fact that $\Delta S=0$ for Raman excitations.

\begin{figure}
\begin{centering}
\includegraphics[width=0.8\columnwidth]{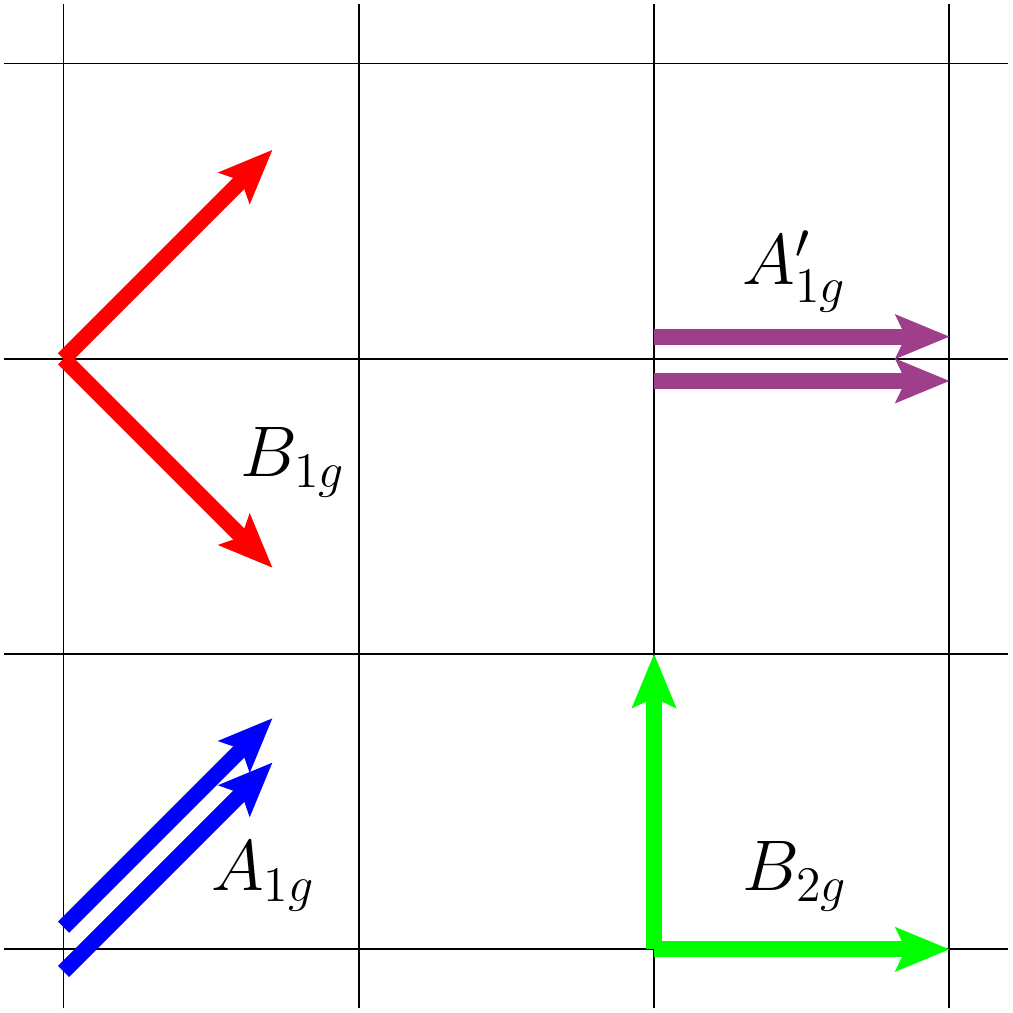}
\par\end{centering}

\centering{}\caption{Polarization geometries for the incident and scattered photons. The
present decomposition refers to the scattering sectors at second order
in the derivation. Higher order terms will mix the sectors as indicated
in Table I of Ref.~\onlinecite{shastry90}.}

\label{fig:Polarization-geometries}
\end{figure}

The effective scattering Hamiltonian obviously depends on the chosen
polarizations of the incoming and scattered photons; to clarify the
discussion we present in Fig.\ref{fig:Polarization-geometries} the
different geometries we have investigated.

The scattering operator can in general be decomposed in two different
ways: the first one involves the symmetry of this operator and
the second refers to the different incoming and outgoing polarization
vectors. These two decompositions are equivalent at second order,
but they are different at higher order (see Ref.~\onlinecite{freericks01}).
Here, we choose to decompose the scattering operator with respect
to the polarization geometries shown in Fig.\ref{fig:Polarization-geometries}.
To stick to the most conventional notations, the top-left crossed
polarization geometry of Fig.\ref{fig:Polarization-geometries} will
be referred to as $B_{1g}$, while the bottom-left geometry will
be called $A_{1g}$.

\subsubsection{Single-band case: results for $S=1/2$}

This calculation has already been done in Ref.~\onlinecite{shastry90}.
However, Ko \emph{et al}.\cite{Lee10} recently pointed out some differences
in some channels between the operator calculated initially and their
own calculation. So, as a warm up, we have rederived the operators
for every channel. We found further differences at fourth order
with respect to the initial calculation.\cite{shastry90,shastry91}
Defining $\mathcal{P}_{\alpha}\equiv\sum_{r}\bm{S}_{r}\cdot\bm{S}_{r+\alpha}$,
$\mathcal{Q}_{\alpha,\beta,\delta}\equiv\sum_{r}(\bm{S}_{r}\cdot\bm{S}_{r+\alpha})(\bm{S}_{r+\beta}\cdot\bm{S}_{r+\delta})$
and $\Delta\equiv t/(U-\omega_{i})$, we found:

\begin{eqnarray}
\mathcal{O}_{B_{1g}} & = & 4t\Delta\left(\frac{1}{2}+2\Delta^{2}\right)\left(\mathcal{P}_{y}-\mathcal{P}_{x}\right)-2t\Delta^{3}\left(\mathcal{P}_{2y}-\mathcal{P}_{2x}\right)\nonumber \\[5mm]
\mathcal{O}_{A_{1g}} & = & 4t\Delta\left(\frac{1}{2}+6\Delta^{2}\right)\left(\mathcal{P}_{y}+\mathcal{P}_{x}\right)-2t\Delta^{3}\left(\mathcal{P}_{2y}+\mathcal{P}_{2x}\right)\nonumber \\
 & - & 8t\Delta^{3}\left(\mathcal{P}_{x+y}+\mathcal{P}_{x-y}\right)\nonumber \\
 & + & 32t\Delta^{3}\left(\mathcal{Q}_{x,y,x+y}+\mathcal{Q}_{y,x,x+y}-\mathcal{Q}_{x+y,x,y}\right)\nonumber \\[5mm]
\mathcal{O}_{B_{2g}} & = & 0\nonumber \\[5mm]
\mathcal{O}_{A'_{1g}} & = & 4t\Delta\left(1+8\Delta^{2}\right)\mathcal{P}_{x}+16t\Delta^{3}\mathcal{P}_{y}\nonumber \\
 & - & 4t\Delta^{3}\mathcal{P}_{2x}-8t\Delta^{3}\left(\mathcal{P}_{x+y}+\mathcal{P}_{x-y}\right)\nonumber \\
 & + & 32t\Delta^{3}\left(\mathcal{Q}_{x,y,x+y}+\mathcal{Q}_{y,x,x+y}-\mathcal{Q}_{x+y,x,y}\right)
\end{eqnarray}

In Appendix \ref{sec:Comparison-of-effective}, we discuss the differences between this operator and
the result of Ref.\onlinecite{shastry90,shastry91} after rewriting it following the
symmetry-based decomposition used in this reference. To test the validity of our operators, we have also compared on small clusters the Raman spectra
of the original single-band Hubbard model with those obtained with the effective operators,
with the conclusion that the spectra obtained with our operators agree much better.

\subsubsection{Two-band case: results for $S=1$}

The general procedure is similar to the one of the single-band case,
but starting from the two-band Hamiltonian described in Subsection
\ref{sub:Microscopic-parameters}. There are two main differences
between this case and the single-band case.

First, in the one-band case, the only excited states that were considered
were the terms with two particles on one site, inducing a Coulomb
repulsion $U$. These terms will still be present in the two-band
case, but other terms where only the Hund's coupling is not satisfied
will also be present. At fourth order, this leads to resonances at
different incoming energies.

Moreover, since spin-1 live in a 3-dimensional space, there are eight non-trivial
hermitian operators acting on-site (instead of three in the spin-1/2 case). We thus expect
new types of operator to appear, as for example the biquadratic coupling $(\bm{S}_{i}\cdot\bm{S}_{j})^{2}$.

To express the Raman operator in this case, we take the same definition
for $\mathcal{P}_{\alpha}$ and $\mathcal{Q}_{\alpha,\beta,\delta}$ as for spin-1/2
and we introduce in addtion the operators $\mathcal{R}_{\alpha}\equiv\sum_{r}(\bm{S}_{r}\cdot\bm{S}_{r+\alpha})^{2}$ and
$\mathcal{T}_{\alpha,\beta}\equiv\sum_{r}(\bm{S}_{r}\cdot\bm{S}_{r+\alpha})(\bm{S}_{r+\alpha}\cdot\bm{S}_{r+\beta})$.
To fourth order in perturbation theory, the resulting operators for $B_{1g}$ and $A_{1g}$ read:

\begin{eqnarray}
\mathcal{O}_{B_{1g}} & = & B_{h}\left(\mathcal{P}_{x}-\mathcal{P}_{y}\right)+B_{b}\left(\mathcal{R}_{x}-\mathcal{R}_{y}\right)\nonumber \\
 & + & B_{h2}\left(\mathcal{P}_{i+2x}-\mathcal{P}_{i+2y}\right)+B_{3}\left(\mathcal{T}_{x,2x}-\mathcal{T}_{y,2y}\right)\nonumber \\[5mm]
\mathcal{O}_{A_{1g}} & = & A_{h}\left(\mathcal{P}_{x}+\mathcal{P}_{y}\right)+A_{b}\left(\mathcal{R}_{x}+\mathcal{R}_{y}\right)\nonumber \\
 & + & A_{d}(\mathcal{P}_{x+y}+\mathcal{P}_{x-y})\nonumber \\
 & + & A_{h2}\left(\mathcal{P}_{2x}+\mathcal{P}_{2y}\right)+A_{3}\left(\mathcal{T}_{x,2x}+\mathcal{T}_{y,2y}\right)\nonumber \\
 & + & A_{3d}\left(\mathcal{T}_{x,x+y}+\mathcal{T}_{x,x-y}+\mathcal{T}_{y,x+y}+\mathcal{T}_{-y,x-y}\right)\nonumber \\
 & + & A_{p}\left(\mathcal{Q}_{x,x+y,y}+\mathcal{Q}_{y,x,x+y}-\mathcal{Q}_{x+y,x,y}\right)\label{eq:effective_raman}
\end{eqnarray}
 The different coefficients are functions of the different parameters
of the initial Hamiltonian, Eq.(\ref{eq:microscopic}). They are given
in Appendix \ref{coefficient}.

\subsection{Relevant limits and geometries}

One of the major advantages of Raman scattering lies in the possibility
of using light polarization to select and characterize the excitation
that one is willing to investigate. As the derivation remains very
systematic, so far, we have taken care of all the different polarizations.
At second order, only $A_{1g}$ and $B_{1g}$ geometries have non-vanishing
operators. Therefore, in the rest of the paper we will mainly focus
our attention on these two cases.

In addition to the usual decomposition due to the incoming and outgoing
polarizations, one should also be aware of the information that can
be accessed through a suitable choice of the incoming photon energy.
The Raman operators and
the associated prefactors of Appendix \ref{coefficient} lead to
many terms.
However, since all the prefactors depend
on the incoming photon energy, it is possible to adjust $\omega_{in}$
to get close to a resonance and highlight specific terms of the scattering
operators.

In the limit $\hbar\omega_{in}\lesssim U+2J_{H}\sim U$, the second
order processes will be dominant, and the other terms can be neglected.
Therefore, in this case, we can restrict ourselves to the Fleury-Loudon
Raman operator:

\[
\hbar\omega_{in}\sim U\Rightarrow\left\lbrace \begin{array}{rcl}
\mathcal{O}_{A1g} & \propto & \sum_{r}\left(\bm{S}_{r}\cdot\bm{S}_{r+x}+\bm{S}_{r}\cdot\bm{S}_{r+y}\right)\\
\mathcal{O}_{B1g} & \propto & \sum_{r}\left(\bm{S}_{r}\cdot\bm{S}_{r+x}-\bm{S}_{r}\cdot\bm{S}_{r+y}\right)
\end{array}\right.
\]

Another interesting limit occurs when $\hbar\omega_{in}\sim4U$. In
this case, we will favor processes where there is an intermediate state
with four electrons at the same site. Such processes occur at
fourth order if they involve only two sites. Moreover, as they
can change the local spin by $\Delta S=2$, they have to be related
to the operator $\left(\bm{S}_{i}\cdot\bm{S}_{j}\right)^{2}$. This
explanation is confirmed by the exact coefficient of the Raman operator
given in Appendix \ref{coefficient}. So in this limit, we can consider
that the Raman operator reduces to:

\[
\hbar\omega_{in}\sim4U\Rightarrow\left\lbrace \begin{array}{rcl}
\mathcal{O}_{A1g} & \propto & \sum_{r}\left(\bm{S}_{r}\cdot\bm{S}_{r+x}\right)^{2}+\left(\bm{S}_{r}\cdot\bm{S}_{r+y}\right)^{2}\\
\mathcal{O}_{B1g} & \propto & \sum_{r}\left(\bm{S}_{r}\cdot\bm{S}_{r+x}\right)^{2}-\left(\bm{S}_{r}\cdot\bm{S}_{r+y}\right)^{2}
\end{array}\right.
\]

So, in two different limits, the Raman operator can be written in a
quite simple way for the $B_{1g}$ and the $A_{1g}$ geometries. The
next section will be devoted to an investigation of the Raman spectra calculated from these
four operators in different situations.

\section{Results: expected spectra}

The goal of this section is to look at the Raman response for specific
angles $\theta$ of Eq.(\ref{eq:bilinear-biquadratic}). We do not
mean here to give a quantitative and exhaustive analysis of the expected
spectra but rather a qualitative description of them and, most importantly
of their evolution throughout the different phases. These investigations
are carried out numerically. The presented spectra have been
obtained with Lanczos and the continued fraction\citep{Dagotto94}
for 16-site square lattice cluster with periodic boundary
conditions. This cluster has additional symmetries\citep{Sandvik},
but this is of no consequence for our present purpose, which is to investigate
the overall shape and behavior of the spectra for different phases.

\begin{figure}
\begin{centering}
\includegraphics[width=1\columnwidth]{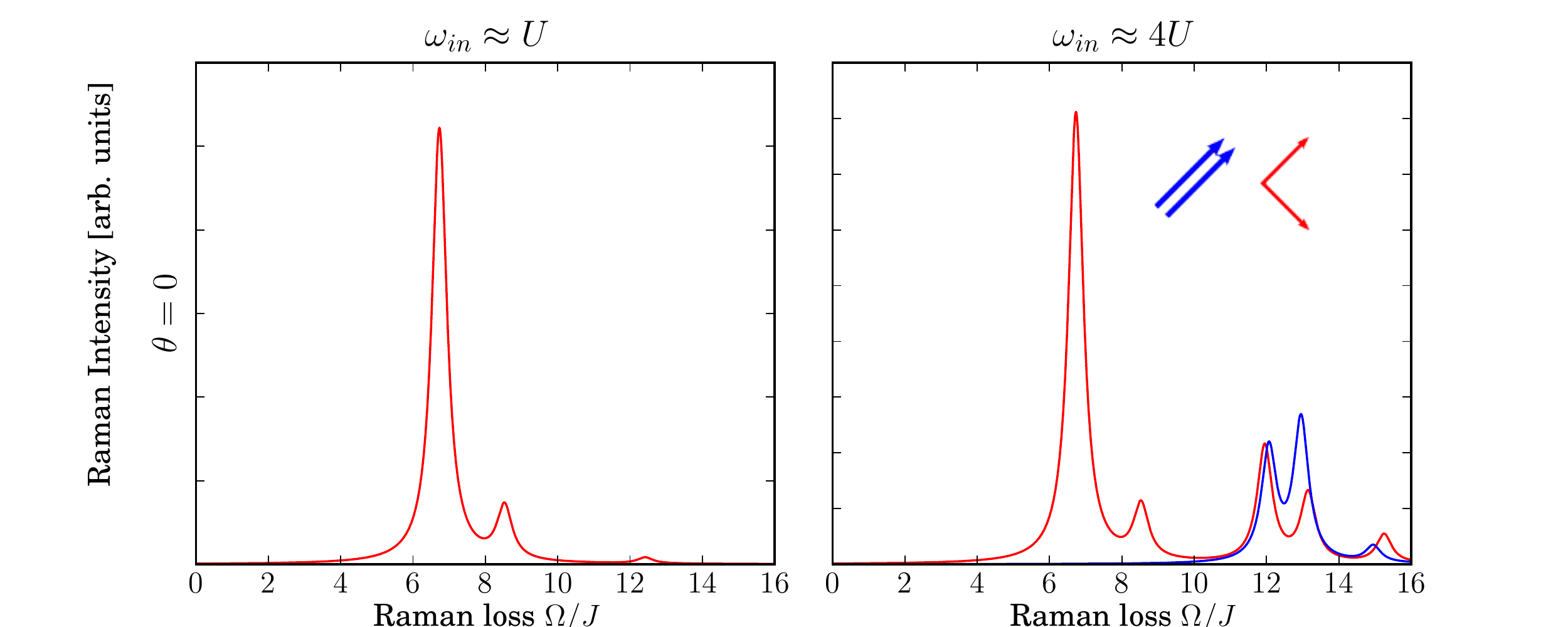}
\par\end{centering}

\centering{}\caption{Raman spectra of the spin-1 Heisenberg model ($\theta=0$ of the Hamiltonian
of Eq. \ref{eq:bilinear-biquadratic}) for $A_{1g}$ and $B_{1g}$ polarizations. The left
column represents the inelastic response as a function of the energy
loss for an incoming photon of energy $\hbar\omega_{in}\lesssim U$,
which corresponds to the Loudon-Fleury operators $\mathcal{O}_{FL}\propto\sum_{i}\left[{\bf S}_{i}\cdot{\bf S}_{i+x}\pm{\bf S}_{i}\cdot{\bf S}_{i+y}\right]$.
The right column corresponds to an incoming photon energy of $\hbar\omega_{in}\lesssim4U$,
and hence to the biquadratic form of the scattering operators $\mathcal{O}_{Biq}\propto\sum_{i}\left[\left({\bf S}_{i}\cdot{\bf S}_{i+x}\right)^{2}\pm\left({\bf S}_{i}\cdot{\bf S}_{i+y}\right)^{2}\right]$.
The spectra have been computed
for a 16-site cluster. \label{fig:Raman-spectra-heis}}
\end{figure}

The results and spectra discussed in the present section are summarized
in Fig.\ref{fig:Raman-spectra-heis} and in Fig.\ref{fig:Raman-spectra-biq}

\subsection{Heisenberg AF phase}

The most conventional Heisenberg antiferromagnetic phase is obviously
reached for $\theta=0$ in the Hamiltonian of Eq.(\ref{eq:bilinear-biquadratic}).
The Raman response in the $B_{1g}$ geometry for the Fleury-Loudon
operator exhibits a bi-magnon peak which can be understood with a
standard spin-wave calculation: Linear spin-wave theory leads to a peak at a Raman loss
$\hbar\Omega=\hbar\left(\omega_{in}-\omega_{out}\right)\approx8J$ which softens down to $\hbar\Omega\approx7.8J$
after taking into account magnon-magnon interactions, as shown for
instance by Chubukov and Frenkel\citep{Chubukov95} or Canali and
Girvin.\citep{Canali-Girvin} This is in qualitative agreement with
our numerical result. For the same polarization, the biquadratic operator
gives the same response with additional peaks for higher Raman losses
corresponding to multi-magnon excitations.

In the $A_{1g}$ geometry, the Fleury-Loudon operator commutes with
the Heisenberg Hamiltonian and the Raman response vanishes. This is
not the case for the biquadratic operator: as in the Fleury-Loudon
case, the two-magnon features disappear, but the multi-magnon peaks,
which arise from the non-commutating part of the operator with the Hamiltonian,
remain visible. It would be really interesting to check this prediction
experimentally, namely that intensity shows up at larger Raman loss energy when the
incoming light frequency is in the range $4U/\hbar$.

\subsection{Ferro-Quadrupolar order}

\begin{figure}
\begin{centering}
\includegraphics[width=1\columnwidth]{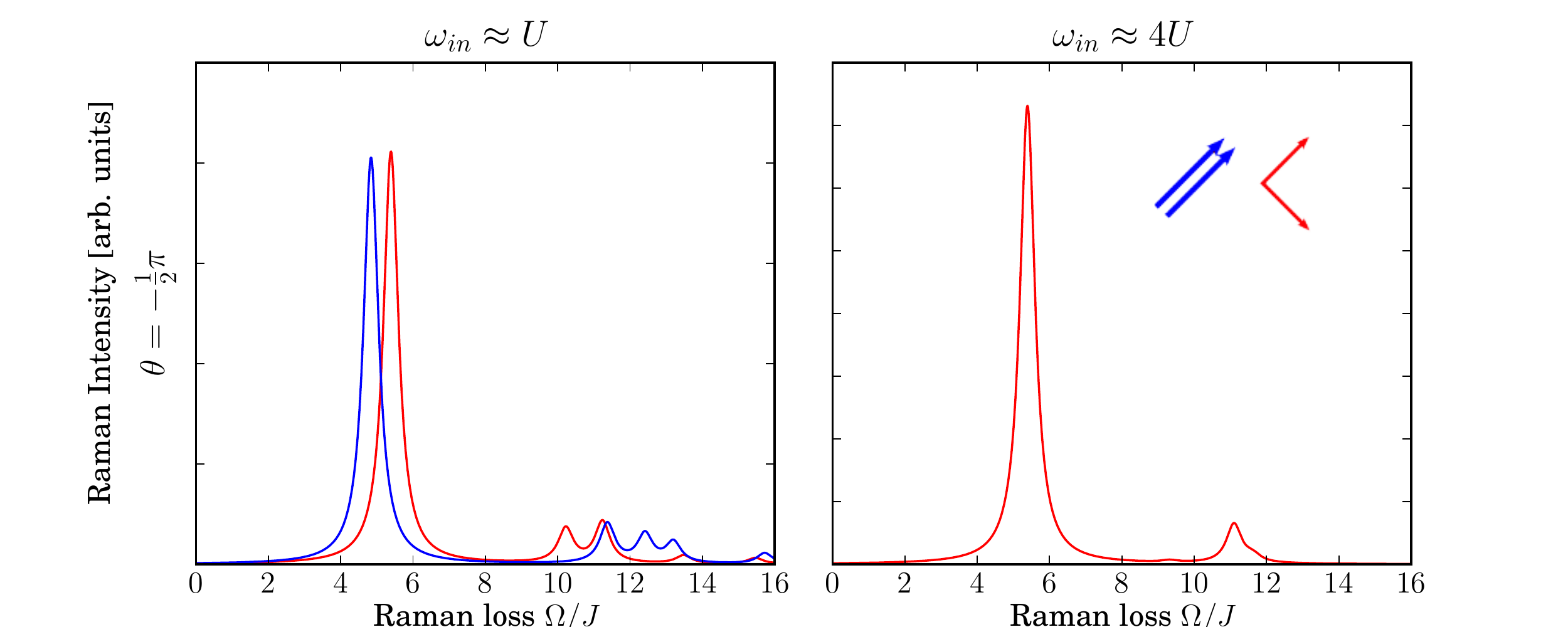}
\par\end{centering}

\centering{}\caption{Raman spectra for a model with purely biquadratic coupling ($\theta=-\frac{\pi}{2}$ of the
Hamiltonian of Eq. \ref{eq:bilinear-biquadratic}) for $A_{1g}$ and $B_{1g}$ polarizations. The left
column represents the inelastic response as a function of the energy
loss for an incoming photon of energy $\hbar\omega_{in}\lesssim U$,
which corresponds to the Loudon-Fleury operators $\mathcal{O}_{FL}\propto\sum_{i}\left[{\bf S}_{i}\cdot{\bf S}_{i+x}\pm{\bf S}_{i}\cdot{\bf S}_{i+y}\right]$.
The right column corresponds to an incoming photon energy of $\hbar\omega_{in}\lesssim4U$,
and hence to the biquadratic form of the scattering operators $\mathcal{O}_{Biq}\propto\sum_{i}\left[\left({\bf S}_{i}\cdot{\bf S}_{i+x}\right)^{2}\pm\left({\bf S}_{i}\cdot{\bf S}_{i+y}\right)^{2}\right]$.
The spectra have
been computed for a 16-site cluster. \label{fig:Raman-spectra-biq}}
\end{figure}

For quadrupolar order,
the relevant order parameter, ${\bf Q}$ (see Ref. ~\onlinecite{Penc_Laeuchli}), and its descendant
observables like ${\bf Q}\cdot{\bf Q}$ can be written in terms of
spin operators:
\begin{equation}
{\bf Q}_{i}\cdot{\bf Q}_{j}=2\left({\bf S}_{i}\cdot{\bf S}_{j}\right)^{2}+{\bf S}_{i}\cdot{\bf S}_{j}+Cst.\label{eq:QQ}
\end{equation}
 An easy point to investigate corresponds to $\theta=-\pi/2$ where
the Hamiltonian reduces to its biquadratic part.

As already discussed, there are two resonant regimes: the first one
for an incoming photon energy $\hbar\omega_{in}\lesssim U$ leading
to the bilinear Fleury-Loudon operator, and a second one for $\hbar\omega_{in}\lesssim4U$
at which the leading term of the scattering operator is proportional
to $\left({\bf S}_{i}\cdot{\bf S}_{j}\right)^{2}$. Since in Eq.(\ref{eq:QQ})
both bilinear and biquadratic couplings appear in the quadrupole-quadrupole observable,
both resonant regimes offer the possibility to investigate quadrupolar
order.

Concerning the physics occuring at $\hbar\omega_{in}\lesssim U$,
the Fleury-Loudon scattering Hamiltonian is expected to give access
to quadrupolar excitations, and this turns out to be the case: The
Raman response exhibits a marked peak for a Raman loss of $\hbar\Omega\approx5J$,
which can be understood as coming from a pair of quadrupolar excitations. Indeed,
a flavor-wave\cite{Papanicolaou,Penc_Laeuchli} calculation gives
the dispersion of the quadrupolar excitations across the Brillouin
zone, as displayed in Fig.\ref{fig:Flavor-wave-dispersion}, with
a $2J$ excitation at $(\pi,0)$, leading to a peak close to a Raman loss of $4J$. 
This prediction corresponds to the
excitation of two non-interacting quadrupolar excitations. However,
this simple picture does not take into account particle-particle
interactions, which in the present case appear to shift the peak position at higher energy. At first glance, it
is not obvious that the Loudon-Fleury scattering operator offers the
possibility to investigate quadrupolar excitations since this scattering
resonance is mostly known to highlight the bi-magnon peak. Yet, in
contrast to the conventional magnetic scattering, the two scattering
geometries, $A_{1g}$ and $B_{1g}$, have comparable spectra, as none
of the operator commutes with the Hamiltonian. The similarity of the two spectra
for the two polarization geometries seems to be a signature of a
ferro-quadrupolar phase.

The second resonance, for an incoming photon energy of $\hbar\omega_{in}\lesssim4U$
, is associated to a scattering operator with biquadratic spin terms.
As quadrupolar ordered phases are ground state of the Hamiltonian
when $\left({\bf S}_{i}\cdot{\bf S}_{j}\right)^{2}$ dominates, one
expects that the effective Raman operator for $\hbar\omega_{in}\lesssim4U$
probes directly quadrupolar excitations. Indeed, this is confirmed
by our calculations and the obtained results can be explained in a
straightforward manner.
For the $A_{1g}$ geometry, the Raman operator commutes with the Hamiltonian,
hence the response vanishes. One should note however that, for real
systems many terms appear in the scattering operator and longer-range
terms in the Hamiltonian should also be considered; all this facts
combine to give a finite Raman response even in this geometry. Yet,
exactly like for conventional magnetic Raman scattering, the amplitude
of the response should be much stronger for $B_{1g}$ polarizations, as
shown in Fig.\ref{fig:Raman-spectra-heis}. One recovers the quadrupolar
excitation at $\hbar\Omega\approx5J$.

\begin{center}
\begin{figure}
\begin{centering}
\includegraphics[width=0.9\columnwidth]{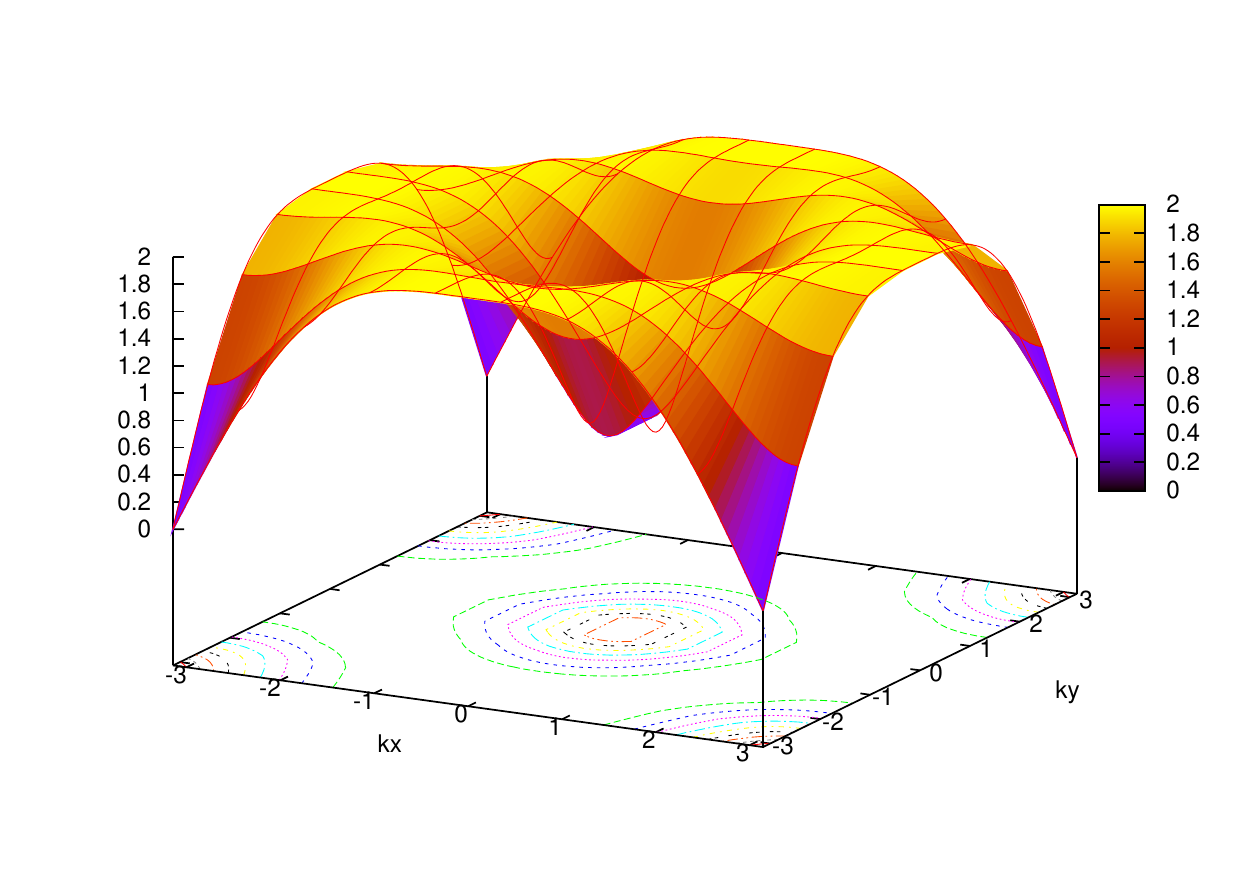}
\par\end{centering}

\caption{Flavor-wave dispersion for a ferro-quadrupolar order corresponding
to $\theta=-\pi/2$ in Eq. (\ref{eq:bilinear-biquadratic}).\label{fig:Flavor-wave-dispersion}}
\end{figure}

\par\end{center}

\section{Discussion}

As summarized in Figs.\ref{fig:Raman-spectra-heis} \& \ref{fig:Raman-spectra-biq},
the signal corresponding to a Néel ordered and a ferro-quadrupolar
phase are qualitatively different. Hence, Raman scattering offers
the opportunity to detect quadrupolar ordering in a relatively straightforward
manner. The systematic is rather simple: one should analyze the Raman
spectra obtained for two scattering geometries (parallel and crossed
polarizations) as presented here and tune the incoming photon energy,
this last step being of course the most crucial one as it enables
one to switch from the standard Loudon-Fleury limit to the biquadratic
form of the scattering operator. This is one of the main messages of
the paper: the microscopic derivation of an effective magnetic scattering
operator for $S=\frac{1}{2}$ systems leads to one resonance at $\hbar\omega_{in}\lesssim U$
since there is only one possible intermediate state. This situation
is no longer valid for $S=1$ compounds, as a spin-1 is formed by
a system of two electrons per site strongly coupled \emph{via} Hund's
rule, different intermediate states can be accessed: one with 3 electrons
at one site, leading to the usual resonance at $\hbar\omega_{in}\lesssim U$;
and another one with 4 electrons at one site leading to a second resonance
at $\hbar\omega_{in}\lesssim4U$. Of course, many other intermediate
states are possible, however, at second order, the only possibility
is to have 3 electrons at one site, therefore the Loudon-Fleury term
will dominate; and, at fourth order biquadratic terms are only appearing
along with intermediate states having quadruple occupancies.

However, going from the first resonant regime ($\hbar\omega_{in}\lesssim U$)
to the second one ($\hbar\omega_{in}\lesssim4U$) requires an adequacy between the
materials and the available light source. Indeed, having a handle on the in and out polarizations
remains much easier for visible light, and if $U$ is too large, one would
end up with incoming photons in the UV region of the light spectrum.
A promising route might be to try organic systems, in which all interactions,
including $U$, are smaller than in oxides.

Turning to the specific case of the ferro-quadrupolar phase, the experimental
investigation would require two steps: i) for an incoming photon energy
of $\hbar\omega_{in}\lesssim U$, one should start with a geometry
corresponding to the $A_{1g}$ polarization and observe a spectrum
with a peak at about $\hbar\Omega\approx5J$. By keeping the same
photon energy and slowly rotating the outgoing polarization until
getting to the $B_{1g}$ polarization, one should not observe much
variations but a small hardening of the main peak. ii) The
second step consists in tuning the incoming photon energy to $\hbar\omega_{in}\lesssim4U$
and to collect the spectra for the different polarizations from $B_{1g}$
to $A_{1g}$, the Raman response should this time exhibit strong modifications:
going from a marked peak to much broader features. In principle, performing
the experiment within the Loudon-Fleury limit should be sufficient.
However, if one does not know precisely the value of $J$ in the considered
compound, this two-step procedure allows a clear identification of
the ferro-quadrupolar order. Precise calculations for the specific
compound would also give information about $J$ after fitting the
data. Also, and from a more fundamental perspective, one should notice
that the second resonance regime should exist even in the case of
a standard Heisenberg phase. In this situation, the Raman spectra
should exhibit not only a bi-magnon peak, but also more spectral weight
at higher energy loss.

\section{Conclusion}

In the present paper, we have derived a general effective inelastic
light-scattering operator for spin-1. We have shown that this operator
offers two different resonant regimes depending on the choice of the
incoming photon energy. On the basis of calculated spectra, obtained
by exact-diagonalization of finite clusters, we have shown that the
different phases of interest (Néel and ferro-quadrupolar) exhibit
characteristic fingerprints that allow a clear identification of each
type of ordering. This work, for the square lattice, illustrates the
potential of Raman scattering technique as a probe for characterizing
quadrupolar order; one of the best candidate for such a phase remains
NiGa$_{2}$S$_{4}$ where the spin $S=1$ located on the Ni$^{2+}$
ions form a triangular lattice. It is clear that a direct application
for this compound is not entirely possible as it requires another
derivation and a suitable decomposition of the scattering channels,
which is left for further investigation.

\section*{Acknowledgments}

We would like to thank H.M. R\o{}nnow and T.A. T\'oth for useful discussions. F.V.
would like to thank EPFL and the CTMC group for hospitality. MaNEP
and SNF are acknowledged for financial support.

\appendix

\section{Comparison of effective Raman operators\label{sec:Comparison-of-effective}}

In this Appendix we aim at comparing an effective magnetic Raman scattering
operator in the Mott insulating state for a spin $S=\frac{1}{2}$
model to the original Raman current operator that is associated to
the Hubbard model. We present here a comparison for the Hubbard model
on an $4\times2=8$-site ladder for $B_{1g}$ polarization as well
as for a 6-site chain. The effective scattering operator derived in
the present work showing some differences compared to Shastry and
Shraiman's, we also provide graphs to compare our results to theirs.

In order to stick to Shastry and Shraiman's original notations of
Ref.~\onlinecite{shastry90}, we remind their results to the reader:

\begin{widetext}

\begin{equation}
\begin{array}{rcl}
\mathcal{O}_{s} & = & 2t\Delta\left(\mathcal{P}_{y}+\mathcal{P}_{x}\right)+8t\Delta^{3}\left(\mathcal{P}_{2x}+\mathcal{P}_{2y}+\mathcal{P}_{x+y}+\mathcal{P}_{x-y}\right)+32t\Delta^{3}\left(\mathcal{Q}_{x,y,x+y}+\mathcal{Q}_{y,x,x+y}-\mathcal{Q}_{x+y,x,y}\right)\\
\\
\mathcal{O}_{d} & = & 4t\Delta\left[\frac{1}{2}-4\Delta^{2}\right]\left(\mathcal{P}_{y}-\mathcal{P}_{x}\right)+8t\Delta^{3}\left(\mathcal{P}_{2y}-\mathcal{P}_{2x}\right)\\
\\
\mathcal{O}_{o} & = & 64t\Delta^{3}\sum_{r}i\epsilon_{\mu,\mu'}{\bf {S}_{r}\cdot\left({\bf {S}_{r+\mu}\times{\bf {S}_{r+\mu'}}}\right)}\\
\\
\mathcal{O}_{e} & = & -16t\Delta^{3}\left(\mathcal{P}_{x+y}+\mathcal{P}_{y-x}\right)
\end{array}\label{shastry}
\end{equation}

While our derivation gives:
\begin{equation}
\begin{array}{rcl}
\mathcal{O}_{s} & = & \left(2t\Delta+24t\Delta^{3}\right)\left(\mathcal{P}_{y}+\mathcal{P}_{x}\right)-2t\Delta^{3}\left(\mathcal{P}_{2x}+\mathcal{P}_{2y}\right)-8t\Delta^{3}\left(\mathcal{P}_{x+y}+\mathcal{P}_{x-y}\right)\\
 & + & 32t\Delta^{3}\left(\mathcal{Q}_{x,y,x+y}+\mathcal{Q}_{y,x,x+y}-\mathcal{Q}_{x+y,x,y}\right)\\
\\
\mathcal{O}_{d} & = & 4t\Delta\left[\frac{1}{2}+2\Delta^{2}\right]\left(\mathcal{P}_{y}-\mathcal{P}_{x}\right)-2t\Delta^{3}\left(\mathcal{P}_{2y}-\mathcal{P}_{2x}\right)\\
\\
\mathcal{O}_{o} & = & 0\\
\\
\mathcal{O}_{e} & = & 0
\end{array}\label{our_derivation}
\end{equation}

\end{widetext}

Ko \emph{et al.\cite{Lee10}} already noticed that there is no chiral
term appearing in $\mathcal{O}_{o}$, unlike in Shastry and Shraiman's
derivation. To check our derivation we perform a numerical analysis
on finite clusters. The quantities we compare are the height of a most
prominent peak in the exact model and the height of the same peak
in the effective model at second and fourth order. We performed this
analysis on two different clusters:
\begin{enumerate}
\item A $2\times4=8$-site cluster with open boundary conditions. This
choice is motivated by the fact that it is the smallest cluster where
some bounds have the correct prefactor in the $x$ direction (for
smaller clusters, some fourth-order renormalization factors will no
be present in the term $\bm{S}_{i}\cdot\bm{S}_{i+x}$ ), while periodic
boundary conditions would lead to the renormalization of some terms
with respect to the infinite case
coming from processes where a doublon goes through the four sites of
one leg. The results for this cluster are presented in Fig.{[}\ref{ladder}{]}
\begin{figure}
\begin{centering}
 \includegraphics[width=1\columnwidth]{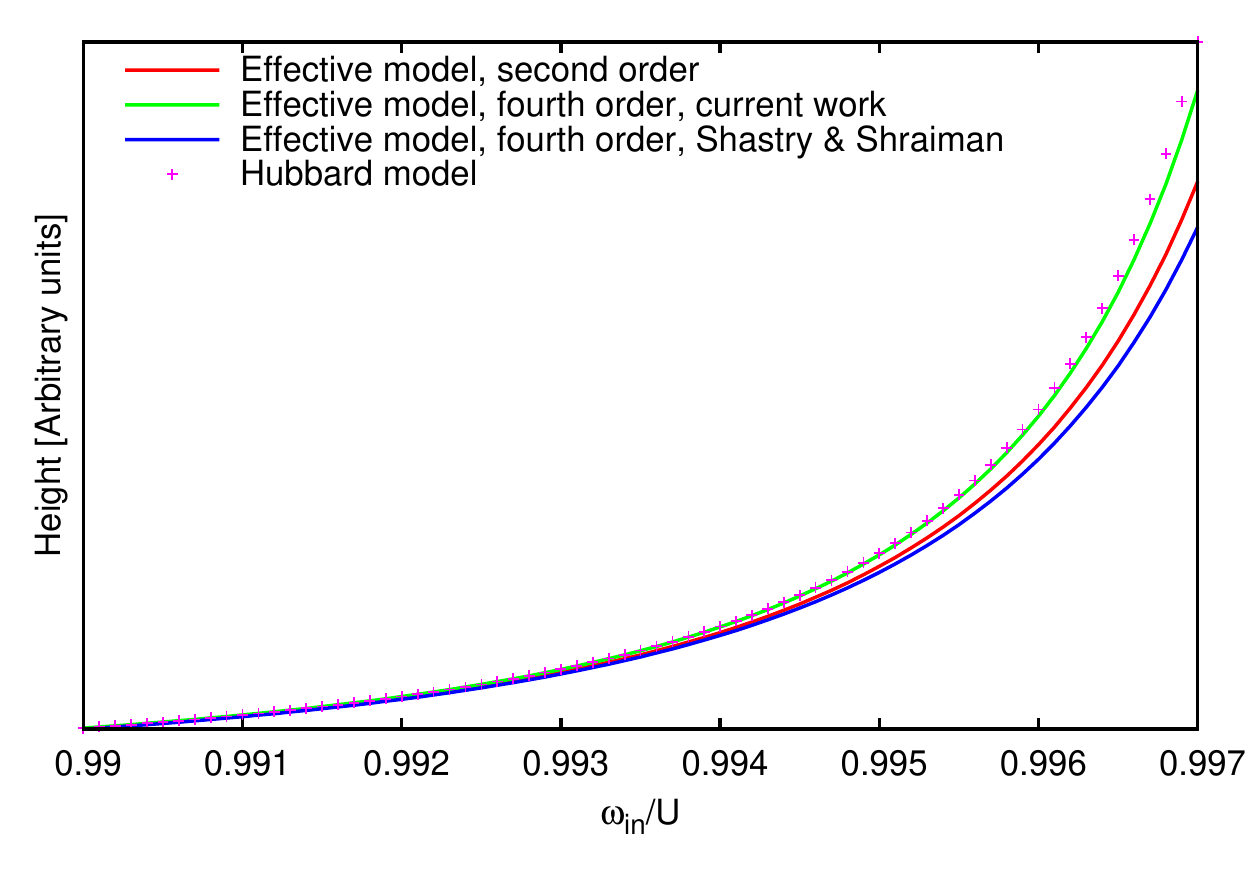}
\par\end{centering}

\caption{Comparison of the height of the main Raman peak for a fixed Raman
loss as a function of the incoming photon energy for different models for a
$2\times4$-site ladder with open boundary conditions. The hopping amplitude has
been set to $t=\frac{U}{2000}$ so that $t\ll U$ and $J=\frac{4t^{2}}{U}=\frac{U}{10^{6}}$}
\label{ladder}
\end{figure}

\item A $6$-site chain with periodic boundary conditions. This
cluster is very usefull to check the prefactor of $\bm{S}_{i}\cdot\bm{S}_{i+2x}$.
As the term $\bm{S}_{i}\cdot\bm{S}_{i+x}$ commutes with the Hamiltonian
on this cluster, only the $\bm{S}_{i}\cdot\bm{S}_{i+2x}$ term contributes
to the scattering amplitude, thus we can check very precisely the
coefficient. Here, it is not possible for a doublon to travel through
the entire system at fourth order, and therefore we can choose periodic
boundary conditions. The results for this cluster are presented in
Fig.{[}\ref{chain}{]}

\begin{figure}
\begin{centering}
 \includegraphics[width=1\columnwidth]{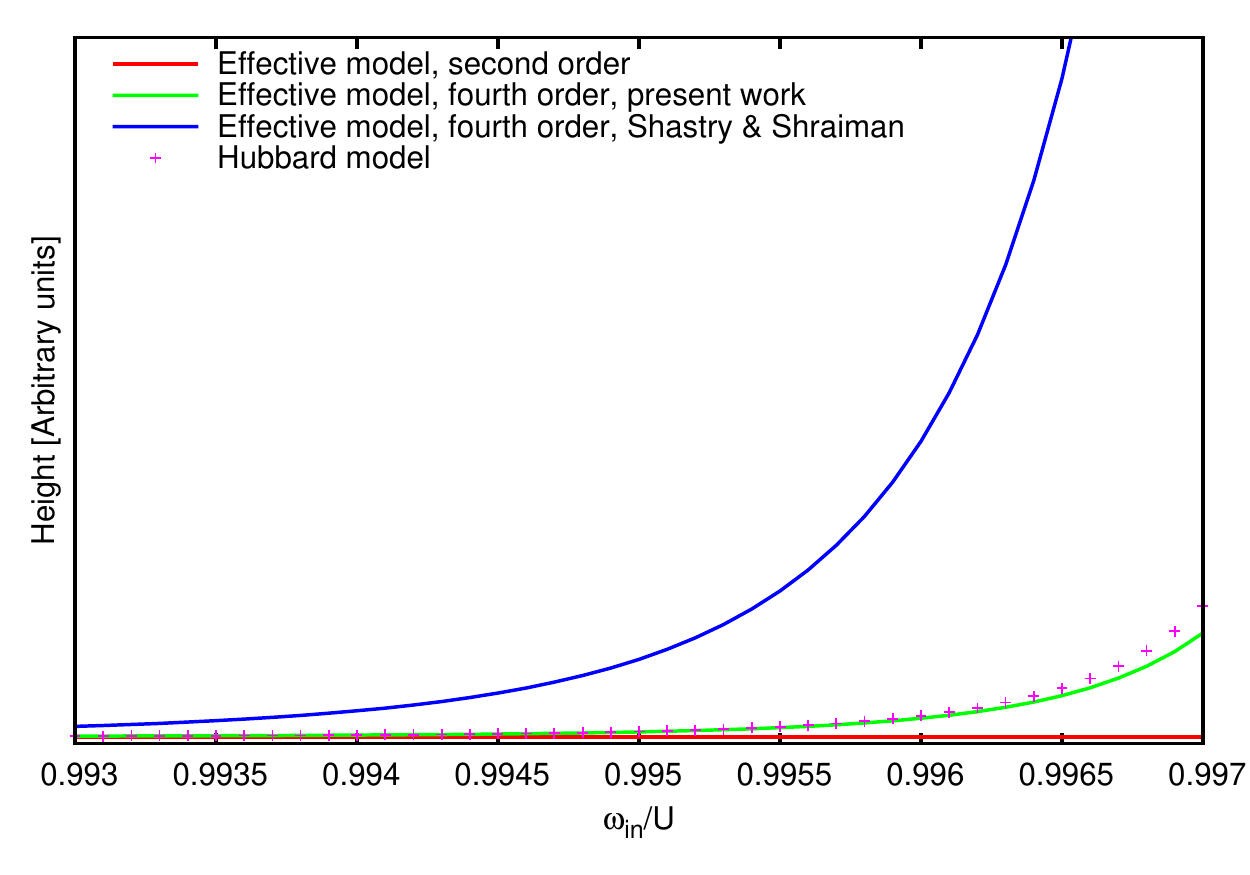}
\par\end{centering}

\caption{Same as Fig. [\ref{ladder}] for a $6$-site chain
with periodic boundary conditions.}
\label{chain}
\end{figure}

\end{enumerate}

As can be seen in these figures, in both cases the spectrum derived from our effective
Raman operator agrees very well with that calculated directly with the original Hubbard
model, while the effective operator of Shastry and Shraiman leads to significant differences
beyond the second order. 

\section{Coefficients of the effective spin-1 Raman operator}

\label{coefficient}

\begin{widetext}

In this Appendix, we list the expression of the prefactors of the effective
spin-1 Raman operator defined in Eq.(\ref{eq:effective_raman}). The terms enclosed
in a square box are those that dominate at the two resonances. 

\begin{eqnarray*}
B_{h} & = & \boxed{\frac{\text{t}_{\text{aa}}^{2}+2\text{t}_{\text{ab}}^{2}+\text{t}_{\text{bb}}^{2}}{2\left(2\text{J}_{H}+U-\omega_{\text{in}}\right)}}+\frac{2\text{t}_{1}^{4}+8\text{t}_{2}^{4}+4\text{t}_{4}^{4}+2\text{t}_{\text{ab}}^{4}}{\left(2\text{J}_{H}+U-\omega_{\text{in}}\right){}^{3}}\\
 & + & \frac{2\left(\text{t}_{1}^{4}-2\text{t}_{4}^{4}+\text{t}_{\text{ab}}^{4}\right)}{\left(2\text{J}_{H}+U-\omega_{\text{in}}\right){}^{2}\left(4\text{J}_{H}+U-\omega_{\text{in}}\right)}-\frac{4\text{t}_{4}^{4}}{\left(2\text{J}_{H}+U-\omega_{\text{in}}\right){}^{2}\left(5\text{J}_{H}-2\text{J}_{P}+U-\omega_{\text{in}}\right)}\\
 & + & \frac{2\left(\text{t}_{2p}^{4}+\text{t}_{\text{ab}}^{4}\right)}{\left(2\text{J}_{H}+U-\omega_{\text{in}}\right){}^{2}\left(5\text{J}_{H}-2\text{J}_{P}+U-\omega_{\text{in}}\right)}+\frac{2\left(4\text{t}_{2}^{4}+\text{t}_{2p}^{4}+2\text{t}_{4}^{4}+\text{t}_{\text{ab}}^{4}\right)}{\left(2\text{J}_{H}+U-\omega_{\text{in}}\right){}^{2}\left(5\text{J}_{H}+2\text{J}_{P}+U-\omega_{\text{in}}\right)}\\
B_{h2} & = & \frac{-2\text{t}_{1}^{4}-8\text{t}_{2}^{4}-4\text{t}_{4}^{4}-2\text{t}_{\text{ab}}^{4}}{\left(2\text{J}_{H}+U-\omega_{\text{in}}\right){}^{3}}+\frac{\text{t}_{1}^{4}-2\text{t}_{4}^{4}+\text{t}_{\text{ab}}^{4}}{\left(2\text{J}_{H}+U-\omega_{\text{in}}\right){}^{2}\left(4\text{J}_{H}+U-\omega_{\text{in}}\right)}\\
 & + & \frac{\text{t}_{2p}^{4}-2\text{t}_{4}^{4}+\text{t}_{\text{ab}}^{4}}{\left(2\text{J}_{H}+U-\omega_{\text{in}}\right){}^{2}\left(5\text{J}_{H}-2\text{J}_{P}+U-\omega_{\text{in}}\right)}+\frac{4\text{t}_{2}^{4}+\text{t}_{2p}^{4}+2\text{t}_{4}^{4}+\text{t}_{\text{ab}}^{4}}{\left(2\text{J}_{H}+U-\omega_{\text{in}}\right){}^{2}\left(5\text{J}_{H}+2\text{J}_{P}+U-\omega_{\text{in}}\right)}\\
B_{b} & = & \boxed{-\frac{4\left(\text{t}_{2p}^{4}-2\text{t}_{4}^{4}+\text{t}_{\text{ab}}^{4}\right)}{\left(2\text{J}_{H}+U-\omega_{\text{in}}\right){}^{2}\left(4U-\omega_{\text{in}}\right)}}\\
B_{3} & = & \frac{\text{t}_{1}^{4}+4\text{t}_{2}^{4}+2\text{t}_{4}^{4}+\text{t}_{\text{ab}}^{4}}{\left(2\text{J}_{H}+U-\omega_{\text{in}}\right){}^{3}}+\frac{-\text{t}_{1}^{4}+2\text{t}_{4}^{4}-\text{t}_{\text{ab}}^{4}}{\left(2\text{J}_{H}+U-\omega_{\text{in}}\right){}^{2}\left(4\text{J}_{H}+U-\omega_{\text{in}}\right)}\\
 & + & \frac{-\text{t}_{2p}^{4}+2\text{t}_{4}^{4}-\text{t}_{\text{ab}}^{4}}{\left(2\text{J}_{H}+U-\omega_{\text{in}}\right){}^{2}\left(5\text{J}_{H}-2\text{J}_{P}+U-\omega_{\text{in}}\right)}+\frac{-4\text{t}_{2}^{4}-\text{t}_{2p}^{4}-2\text{t}_{4}^{4}-\text{t}_{\text{ab}}^{4}}{\left(2\text{J}_{H}+U-\omega_{\text{in}}\right){}^{2}\left(5\text{J}_{H}+2\text{J}_{P}+U-\omega_{\text{in}}\right)}
\end{eqnarray*}

where the parameters $t_{1}$, $t_{2}$, $t_{2p}$, $t_{4}$ are defined
in term of the original microscopic hopping parameters by:
\begin{eqnarray*}
\text{t}_{\text{1}}^{4} & = & \frac{1}{2}\left(\text{t}_{\text{aa}}^{4}+\text{t}_{\text{bb}}^{4}\right)\\
\text{t}_{\text{2}}^{4} & = & \frac{1}{2}\left(\text{t}_{\text{aa}}^{2}\text{t}_{\text{ab}}^{2}+\text{t}_{\text{ab}}^{2}\text{t}_{\text{bb}}^{2}\right)\\
\text{t}_{\text{2p}}^{4} & = & \text{t}_{\text{aa}}^{2}\,\text{t}_{\text{bb}}^{2}\\
\text{t}_{\text{4}}^{4} & = & \text{t}_{\text{aa}}\,\text{t}_{\text{ab}}^{2}\,\text{t}_{\text{bb}}
\end{eqnarray*}

\begin{eqnarray*}
A_{h} & = & \boxed{\frac{\text{t}_{\text{aa}}^{2}+2\text{t}_{\text{ab}}^{2}+\text{t}_{\text{bb}}^{2}}{2\left(2\text{J}_{H}+U-\omega_{\text{in}}\right)}}+\frac{4\text{t}_{1}^{4}+16\text{t}_{2}^{4}+8\text{t}_{4}^{4}+4\text{t}_{\text{ab}}^{4}}{\left(2\text{J}_{H}+U-\omega_{\text{in}}\right){}^{3}}+\frac{4\left(\text{t}_{1}^{4}-2\text{t}_{4}^{4}+\text{t}_{\text{ab}}^{4}\right)}{\left(2\text{J}_{H}+U-\omega_{\text{in}}\right){}^{2}\left(4\text{J}_{H}+U-\omega_{\text{in}}\right)}\\
 & - & \frac{8\text{t}_{4}^{4}}{\left(2\text{J}_{H}+U-\omega_{\text{in}}\right){}^{2}\left(5\text{J}_{H}-2\text{J}_{P}+U-\omega_{\text{in}}\right)}+\frac{4\left(\text{t}_{2p}^{4}+\text{t}_{\text{ab}}^{4}\right)}{\left(2\text{J}_{H}+U-\omega_{\text{in}}\right){}^{2}\left(5\text{J}_{H}-2\text{J}_{P}+U-\omega_{\text{in}}\right)}\\
 & + & \frac{4\left(4\text{t}_{2}^{4}+\text{t}_{2p}^{4}+2\text{t}_{4}^{4}+\text{t}_{\text{ab}}^{4}\right)}{\left(2\text{J}_{H}+U-\omega_{\text{in}}\right){}^{2}\left(5\text{J}_{H}+2\text{J}_{P}+U-\omega_{\text{in}}\right)}+\frac{4\left(\text{t}_{2p}^{4}-2\text{t}_{4}^{4}+\text{t}_{\text{ab}}^{4}\right)}{\left(2\text{J}_{H}+U-\omega_{\text{in}}\right){}^{2}\left(2\text{J}_{H}+3U-\omega_{\text{in}}\right)}\\[5mm]
A_{h2} & = & \frac{-2\text{t}_{1}^{4}-8\text{t}_{2}^{4}-4\text{t}_{4}^{4}-2\text{t}_{\text{ab}}^{4}}{\left(2\text{J}_{H}+U-\omega_{\text{in}}\right){}^{3}}+\frac{\text{t}_{1}^{4}-2\text{t}_{4}^{4}+\text{t}_{\text{ab}}^{4}}{\left(2\text{J}_{H}+U-\omega_{\text{in}}\right){}^{2}\left(4\text{J}_{H}+U-\omega_{\text{in}}\right)}\\
 & + & \frac{\text{t}_{2p}^{4}-2\text{t}_{4}^{4}+\text{t}_{\text{ab}}^{4}}{\left(2\text{J}_{H}+U-\omega_{\text{in}}\right){}^{2}\left(5\text{J}_{H}-2\text{J}_{P}+U-\omega_{\text{in}}\right)}+\frac{4\text{t}_{2}^{4}+\text{t}_{2p}^{4}+2\text{t}_{4}^{4}+\text{t}_{\text{ab}}^{4}}{\left(2\text{J}_{H}+U-\omega_{\text{in}}\right){}^{2}\left(5\text{J}_{H}+2\text{J}_{P}+U-\omega_{\text{in}}\right)}\\[5mm]
A_{b} & = & \boxed{-\frac{4\left(\text{t}_{2p}^{4}-2\text{t}_{4}^{4}+\text{t}_{\text{ab}}^{4}\right)}{\left(2\text{J}_{H}+U-\omega_{\text{in}}\right){}^{2}\left(4U-\omega_{\text{in}}\right)}}\\[5mm]
A_{3} & = & \frac{\text{t}_{1}^{4}+4\text{t}_{2}^{4}+2\text{t}_{4}^{4}+\text{t}_{\text{ab}}^{4}}{\left(2\text{J}_{H}+U-\omega_{\text{in}}\right){}^{3}}+\frac{-\text{t}_{1}^{4}+2\text{t}_{4}^{4}-\text{t}_{\text{ab}}^{4}}{\left(2\text{J}_{H}+U-\omega_{\text{in}}\right){}^{2}\left(4\text{J}_{H}+U-\omega_{\text{in}}\right)}\\
 & + & \frac{-\text{t}_{2p}^{4}+2\text{t}_{4}^{4}-\text{t}_{\text{ab}}^{4}}{\left(2\text{J}_{H}+U-\omega_{\text{in}}\right){}^{2}\left(5\text{J}_{H}-2\text{J}_{P}+U-\omega_{\text{in}}\right)}+\frac{-4\text{t}_{2}^{4}-\text{t}_{2p}^{4}-2\text{t}_{4}^{4}-\text{t}_{\text{ab}}^{4}}{\left(2\text{J}_{H}+U-\omega_{\text{in}}\right){}^{2}\left(5\text{J}_{H}+2\text{J}_{P}+U-\omega_{\text{in}}\right)}\\[5mm]
A_{d} & = & \frac{-8\text{t}_{1}^{4}-32\text{t}_{2}^{4}-16\text{t}_{4}^{4}-8\text{t}_{\text{ab}}^{4}}{\left(2\text{J}_{H}+U-\omega_{\text{in}}\right){}^{3}}+\frac{2\left(\text{t}_{1}^{4}-2\text{t}_{4}^{4}+\text{t}_{\text{ab}}^{4}\right)}{\left(2\text{J}_{H}+U-\omega_{\text{in}}\right){}^{2}\left(4\text{J}_{H}+U-\omega_{\text{in}}\right)}\\
 & - & \frac{4\text{t}_{4}^{4}}{\left(2\text{J}_{H}+U-\omega_{\text{in}}\right){}^{2}\left(5\text{J}_{H}-2\text{J}_{P}+U-\omega_{\text{in}}\right)}+\frac{2\left(\text{t}_{2p}^{4}+\text{t}_{\text{ab}}^{4}\right)}{\left(2\text{J}_{H}+U-\omega_{\text{in}}\right){}^{2}\left(5\text{J}_{H}-2\text{J}_{P}+U-\omega_{\text{in}}\right)}\\
 & + & \frac{2\left(4\text{t}_{2}^{4}+\text{t}_{2p}^{4}+2\text{t}_{4}^{4}+\text{t}_{\text{ab}}^{4}\right)}{\left(2\text{J}_{H}+U-\omega_{\text{in}}\right){}^{2}\left(5\text{J}_{H}+2\text{J}_{P}+U-\omega_{\text{in}}\right)}\\[5mm]
A_{p} & = & -\frac{2\left(2\text{t}_{1}^{4}+8\text{t}_{2}^{4}+4\text{t}_{4}^{4}+2\text{t}_{\text{ab}}^{4}\right)}{\left(2\text{J}_{H}+U-\omega_{\text{in}}\right){}^{3}}\\
A_{3d} & = & \frac{\text{t}_{1}^{4}+4\text{t}_{2}^{4}+2\text{t}_{4}^{4}+\text{t}_{\text{ab}}^{4}}{\left(2\text{J}_{H}+U-\omega_{\text{in}}\right){}^{3}}+\frac{-\text{t}_{1}^{4}+2\text{t}_{4}^{4}-\text{t}_{\text{ab}}^{4}}{\left(2\text{J}_{H}+U-\omega_{\text{in}}\right){}^{2}\left(4\text{J}_{H}+U-\omega_{\text{in}}\right)}\\
 & + & \frac{-\text{t}_{2p}^{4}+2\text{t}_{4}^{4}-\text{t}_{\text{ab}}^{4}}{\left(2\text{J}_{H}+U-\omega_{\text{in}}\right){}^{2}\left(5\text{J}_{H}-2\text{J}_{P}+U-\omega_{\text{in}}\right)}+\frac{-4\text{t}_{2}^{4}-\text{t}_{2p}^{4}-2\text{t}_{4}^{4}-\text{t}_{\text{ab}}^{4}}{\left(2\text{J}_{H}+U-\omega_{\text{in}}\right){}^{2}\left(5\text{J}_{H}+2\text{J}_{P}+U-\omega_{\text{in}}\right)}
\end{eqnarray*}

\end{widetext}

\end{document}